\documentclass[conference]{IEEEtran}
\IEEEoverridecommandlockouts
\usepackage[noadjust]{cite}
\usepackage{amsmath,amssymb,amsfonts}
\usepackage{algorithm}
\usepackage{algorithmic}
\usepackage{graphicx}
\usepackage{textcomp}
\usepackage{xcolor}
\usepackage{comment}
\usepackage{subfigure}
\usepackage{flushend}

\def\BibTeX{{\rm B\kern-.05em{\sc i\kern-.025em b}\kern-.08em
    T\kern-.1667em\lower.7ex\hbox{E}\kern-.125emX}}
\begin{document}

\title{Boosting the Search Performance of B+-tree for Non-volatile Memory with Sentinels}

\author{\IEEEauthorblockN{Chongnan Ye}
\IEEEauthorblockA{\textit{School of Information Science and Technology}\\
\textit{ShanghaiTech University}\\
yechn@shanghaitech.edu.cn
}
\and
\IEEEauthorblockN{Chundong Wang*\thanks{*Corresponding Author}}
\IEEEauthorblockA{\textit{School of Information Science and Technology}\\ 
\textit{ShanghaiTech University}\\
wangchd@shanghaitech.edu.cn
}
}

\maketitle

\begin{abstract}
The next-generation non-volatile memory (NVM) is striding into computer systems as a new tier 
as it incorporates both DRAM's byte-addressability and disk's persistency.
Researchers and practitioners have considered building {\em persistent memory} 
by placing NVM on the memory bus for CPU to directly load and store data.
As a result, cache-friendly data structures have been developed for NVM.
One of them is the prevalent B+-tree. State-of-the-art in-NVM B+-trees
mainly focus on the optimization of write operations (insertion and deletion).
However, search is of vital importance for B+-tree.
Not only search-intensive workloads benefit from an optimized search, but
insertion and deletion also rely on a preceding search operation to proceed.
In this paper, we attentively study a sorted B+-tree node that spans over
contiguous cache lines. 
Such cache lines
exhibit a monotonically increasing trend and searching a target key across them
can be accelerated by estimating a range the key falls into.
To do so, we construct a probing {\em Sentinel Array} in which a sentinel stands for
each cache line of B+-tree node. Checking the Sentinel Array 
avoids scanning unnecessary cache lines and hence
significantly reduces cache misses for a search.
A quantitative evaluation shows that
using Sentinel Arrays boosts the search performance of state-of-the-art in-NVM B+-trees
by up to 48.4\% while the cost of maintaining of Sentinel Array is low.
\end{abstract}

\begin{IEEEkeywords}
Non-volatile memory, B+-tree, Key-value Database, Cache
\end{IEEEkeywords}

\section{Introduction}
A conventional computer employs DRAM as main memory and hard disk as storage. 
The disk provides persistency but demands a long time to load and store data~\cite{b1}. 
DRAM is volatile and cannot reach large capacity because of its density limitation~\cite{lee2009architecting, qureshi2009scalable, zhou2009durable}. 
Recently, the next-generation non-volatile memory (NVM), such as phase-change memory (PCM)~\cite{raoux2008phase}, spin-transfer torque memory (STT-RAM)~\cite{kawahara2010scalable}, and resistive RAM (RRAM)~\cite{chang2016resistance}, emerges as a promising medium that combines both DRAM's byte-addressability and disk's persistency. Hence, researchers and practitioners have proposed to place NVM on the memory bus to build {\em persistent memory}, which acts
as a new tier of computer architecture and blurs the boundary between DRAM and disk.

Multiple classic data structures have been tuned for NVM. Among them 
B+-tree is one that has been widely used for databases and file systems since the era of disks~
\cite{bayer2002organization}. 
B+-tree is a multi-level structure composed of internal and leaf nodes. 
Internal nodes (INs) are used to index lower-level INs and LNs while leaf nodes (LNs) hold actual key-value (KV) pairs. 
INs and LNs have the same structure but the pointers of INs point to the address of LNs while the pointers of LNs point to the value data that stored in the B+-tree. And INs have one more pointer than LNs which indicates the leftmost LN of the internal nodes.
 Keys are sorted in both INs and LNs. When tuning B+-tree for NVM, 
 researchers need to consider the fact that CPU or memory controller may reorder writes from
 CPU cache to NVM \cite{yang2015nv, hwang2018endurable, wang2019circ}. One way to retain a writing order is to 
employ a combination of cache line flush and memory fence~\cite{moraru2013consistent}, which, however, incurs significant performance overheads. 
With further and further understanding of B+-tree's properties, state-of-the-art 
B+-trees designed for NVM
have evolved from being with sorted nodes (e.g., CDDS-Tree~\cite{venkataraman2011consistent}) to unsorted nodes (e.g., NV-Tree, wB+-tree, and FPTree~\cite{oukid2016fptree}), and again sorted nodes (e.g., FAST-FAIR and Circ-Tree). When developing such B+-tree variants, researchers
 mainly placed their emphasis on reducing the use of cache line flush and memory fence in order to achieve a high write performance for insertion and deletion. 

Nevertheless, the search operation with a target key for the key's corresponding value 
is of paramount importance for B+-tree. 
Search is not only an independent operation, but 
insertion and deletion also rely on a preceding search 
to proceed. 
As the latest FAST-FAIR and Circ-Tree are with sorted nodes, we focus on optimizing the search operation over sorted KV pairs stored in NVM.
It is revealed that linear search is more efficient than binary search over sorted keys although the latter has a theoretically lower time complexity.
The reason is that binary search incurs more cache misses while linear search facilitates CPU's branch prediction and prefetch.
Our aim of this paper is to further reduce cache misses in searching. We find that no matter if we use binary search or linear search,
we must access and check keys that are not the target one. Although accessing those keys leads us to continue the search, it is the cause of cache misses. 
This is the observation of {\em read amplification} in searching a B+-tree node. 
We intend to reduce as much read amplification as possible.
In other words, we want to timely find the exact cache line in which the target key resides. It is impossible except using some auxiliary information.
Previously, Oukid et~al.~\cite{oukid2016fptree} proposed to use 1B hash values as fingerprints for keys. 
A key and its fingerprint have the same offset. Searching a key transforms to finding the key's fingerprint that 
indicates the location of the key. However,
due to the collisions of 1B fingerprints and the calculation cost, fingerprinting is not efficient in practice. 

We closely study a sorted B+-tree node. We find that each cache line has a range of keys and all cache lines
exhibit a monotonically increasing trend. We can choose a {\em sentinel} key in a uniform fashion, e.g., the smallest or the largest key, 
from each cache line to reflect such a trend. When searching a target key, instead of scanning cache lines, we 
go through the Sentinel Array to determine which cache line the target key stays in. 
As a result, searching a target key converts to searching the Sentinel Array and one actual cache line of KV pairs.
The Sentinel Array contains representatives of cache lines of a B+-tree node and it thus
occupies much fewer cache lines. Also, a linear scan of it facilitates CPU's prefetch and branch prediction.
Consequently, employing the Sentinel Array significantly reduces read amplification. 

We have augmented FAST-FAIR and Circ-Tree with
the idea of Sentinel Array. Experimental results show that compared to the use of Sentinel Array boosts the search performance by up to 42.6\%
and 48.4\%, respectively, for FAST-FAIR and Circ-Tree. Moreover, using Sentinel Arrays is up to 60.1\% faster than using the idea of fingerprints.
Maintaining Sentinel Arrays, however, incurs at most 4.1\% and 6.5\% more time for insertions with
FAST-FAIR and Circ-Tree, respectively.

The rest of this paper is organized as follows. The background and motivation are discussed in Section 2. Section 3 shows the detailed design and implementation of Sentinel Array. The evaluation and results are presented in Section 4. In the last, we conclude this paper and look forward to future work.

\section{Background and Motivation}

\subsection{Background}
Byte-addressable non-volatile memory (NVM), such as Phase Change Memory (PCM), Spin-Transfer Torque RAM (STT-RAM), and Resistive RAM (RRAM), has an increasingly important demand in high-performance computing and big data analysis \cite{zhao2013kiln, moraru2013consistent}. It combines the durability of hard disk and the comparable access performance of DRAM. 

Data consistency between the volatile device and the non-volatile device is always one of the most crucial features of storage systems. In the traditional storage architecture, where DRAM acts as the main memory, the system needs to ensure the data consistency from DRAM to the hard disk to prevent the data from being lost if system failure happens \cite{mohan1992aries}. As to NVM, the effort becomes to keep data consistency between CPU and NVM. However, modern CPUs mainly support an atomic write of no more than the memory bus width (8 bytes for 64-bit CPUs) when NVM is put on the memory bus \cite{hwang2018endurable, dulloor2014system, intel64and}. Moreover, compared to the known persistent status of each page in DRAM, the status of data in cache-lines is unknown and the CPU or the memory controller may write the cache-lines back to the main memory in a different order from the programmed order. Make memory writes are in a certain order is critical to the crash consistency of pointer-based data structures. For example, a new B+-tree from a split operation must be written completely before the pointer to it is added to its parental node. The pointer may turn to be dangling if the order is reversed but the crash occurs \cite{yang2015nv} and it's necessary to maintain the writing order. 

Some CPU instructions, such as \textsf{cache line flush} and \textsf{memory fence}, are provided to maintain the written order from cache to memory. \textsf{cache line flush} explicitly invalidates a dirty cache-line and flush it to memory. \textsf{memory fence} makes a barrier that holds back the memory operations after the barrier until those before the barrier complete. However, these instructions incur performance overheads \cite{yang2015nv, hwang2018endurable, lu2014loose}. Therefore, it is necessary to reduce the use of \textsf{cache line flush} and \textsf{memory fence}.
\subsection{Related Work}

Researchers have proposed a number of mechanisms to provide data consistency and optimize the performance of B+-trees developed for NVM. \textsf{FAST-FAIR} \cite{hwang2018endurable} uses two features that CPU only ensures 8-bytes atomic write and duplicate pointers are impossible in B+-tree node. It exploits the store dependencies in shifting a sequence of KV pairs in a node on insertion/deletion. (e.g. $KV_{i} \xrightarrow{} KV_{i+1}$), which reduces the cost of cache line flushes and memory fence. 

\textsf{FAST-FAIR} still follows the classic B+-tree node in a linear structure. \textsf{Circle-Tree}~\cite{wang2019circ} was proposed to reduce write amplification caused by shifting KV pairs in the linear structure. The circular structure supports bidirectional shifting. KV pair is inserted/deleted at the left or right side by deciding which side requires fewer shifting operations. 

Most B+-trees developed for NVM optimize insertion/deletion performance. Although \textsf{FP-Tree}~\cite{oukid2016fptree} use a technique named fingerprinting, which maps the key of KV pair into a byte of fingerprint (hash value), to optimize the search performance by checking the array of fingerprint instead of the original data container. The efficacy of fingerprinting is low due to hash calculations and the collisions of 1B hash values. 

\subsection{Motivation}

\begin{figure}[htbp]
\centerline{\includegraphics[width=0.5\textwidth]{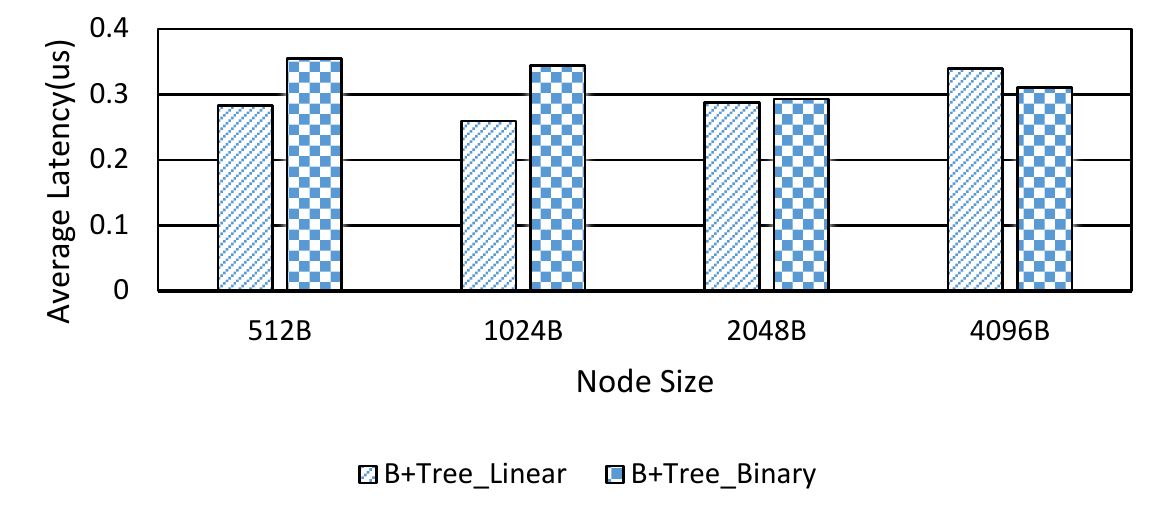}}
\caption{B+-tree with different search way comparison}
\label{L_B_Comparison}
\end{figure}

\begin{figure}[htbp]
\centerline{\includegraphics[width=0.5\textwidth]{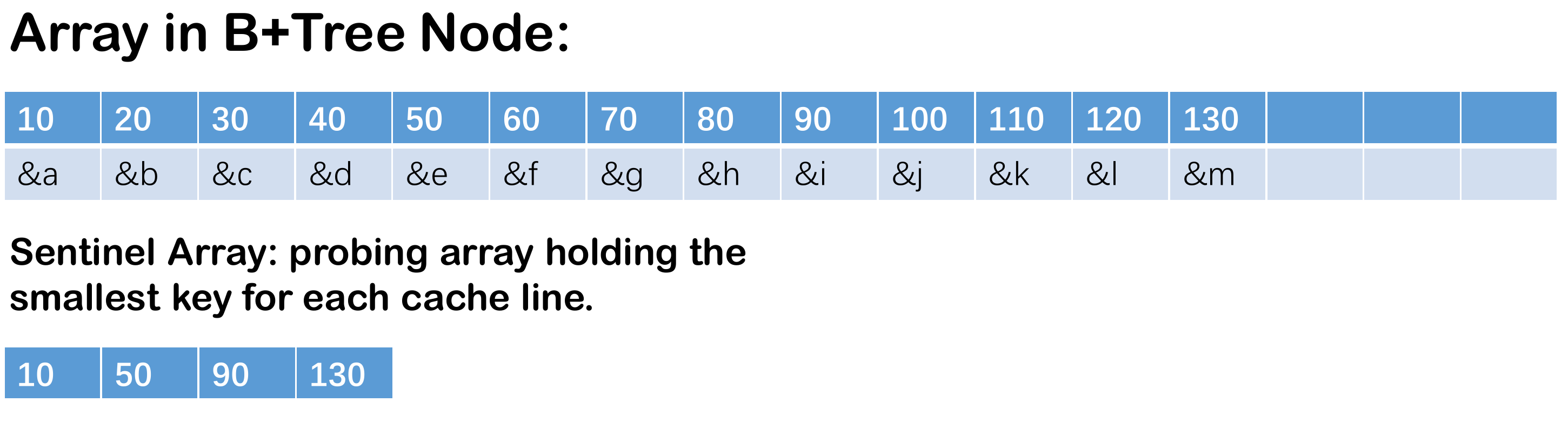}}
\caption{Sentinel Array example}
\label{fig1}
\end{figure}

With further and further understanding of B+-tree's properties, researchers have switched from sorted nodes to unsorted nodes, and again to sorted nodes now. More importantly, the researcher found that a large CPU cache of modern processor makes linear search outperform binary search in a sorted B+-tree node because of the former's good cache line locality \cite{danowitz2012cpu}. Fig.~\ref{L_B_Comparison} verifies that the latency of binary search is greater than that of linear search with small node size (512B/1KB/2KB) when we searched 1 million KV pairs in a sorted B+-tree node. 

As the search is so important that used not only in independent operation, but also in insertion and deletion, optimize the search performance is a key point to boost the throughput. Accessing the key and check if that is the target one is the cause of cache misses. This is the observation of {\em read amplification} in searching a sorted B+-tree node. Our goal focus on reducing cache misses in searching. Fig.~\ref{fig1} shows a sorted B+-tree node with a Sentinel Array which has the smallest key from each cache line. When searching a target key 130, instead of scanning 4 cache lines, we go through the Sentinel Array to determine the 4th cache is the target key stays in. As a result, searching key 130 converts to searching the Sentinel Array and one actual cache line of KV pairs.

\section{Design and Implementation of Sentinel Array}

Sentinel Array is an auxiliary structure that contains sentinels, which indicate the minimum key of each cache line of the origin array. The Sentinel Array size is decided by the machine and the original array size. Given a typical cache line size of 64 bytes and a sentinel (minimum key) in 8B, one cache line contains eight sentinels, which are respectively from eight cache lines of KV pairs. Therefore, a Sentinel Array in one cache line is able to support a B+-tree node in 512B (64$\times$8). For a B+-tree node in 2KB with 32 cache lines, a Sentinel Array with four cache lines suffices. In the worst case for searching over such a 2KB node, all four cache lines of Sentinel Array and one cache line of KV pair are scanned. So in all, there are five cache misses. However, in the worst case of searching the original 2KB node incurs 32 cache misses. Concretely, employing a Sentinel Array significantly reduces cache misses.

We note that Sentinel Array is not enforced consistency with cache line flush and memory fence. Then reason is that Sentinel Array could be reconstructed from keys in the original B+-tree node if the system crashes down. State-of-the-art B+-trees developed for NVM embrace comprehensive mechanisms to guarantee the consistency of KV pairs.

\subsection{Search of Sentinel Array}

\begin{figure}[htbp]
\centerline{\includegraphics[width=0.5\textwidth]{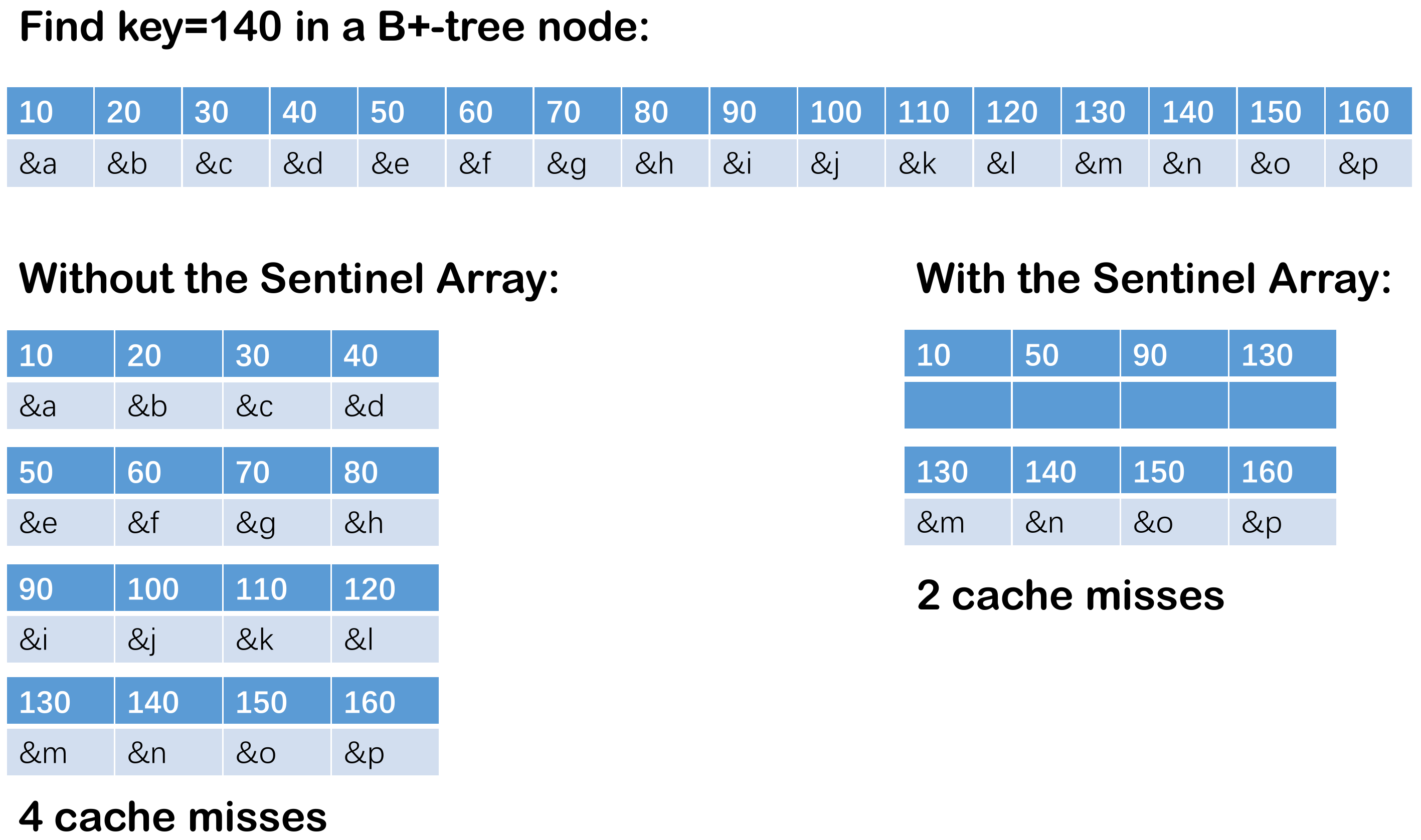}}
\caption{An illustration of searching 140 in a B+-tree node}
\label{search_alg}
\end{figure}

\begin{figure*}[t]
	\centering
	\subfigure[Average latency (1m data, search, $\mu$s)]{
		\centering
		\includegraphics[width=0.48\columnwidth]{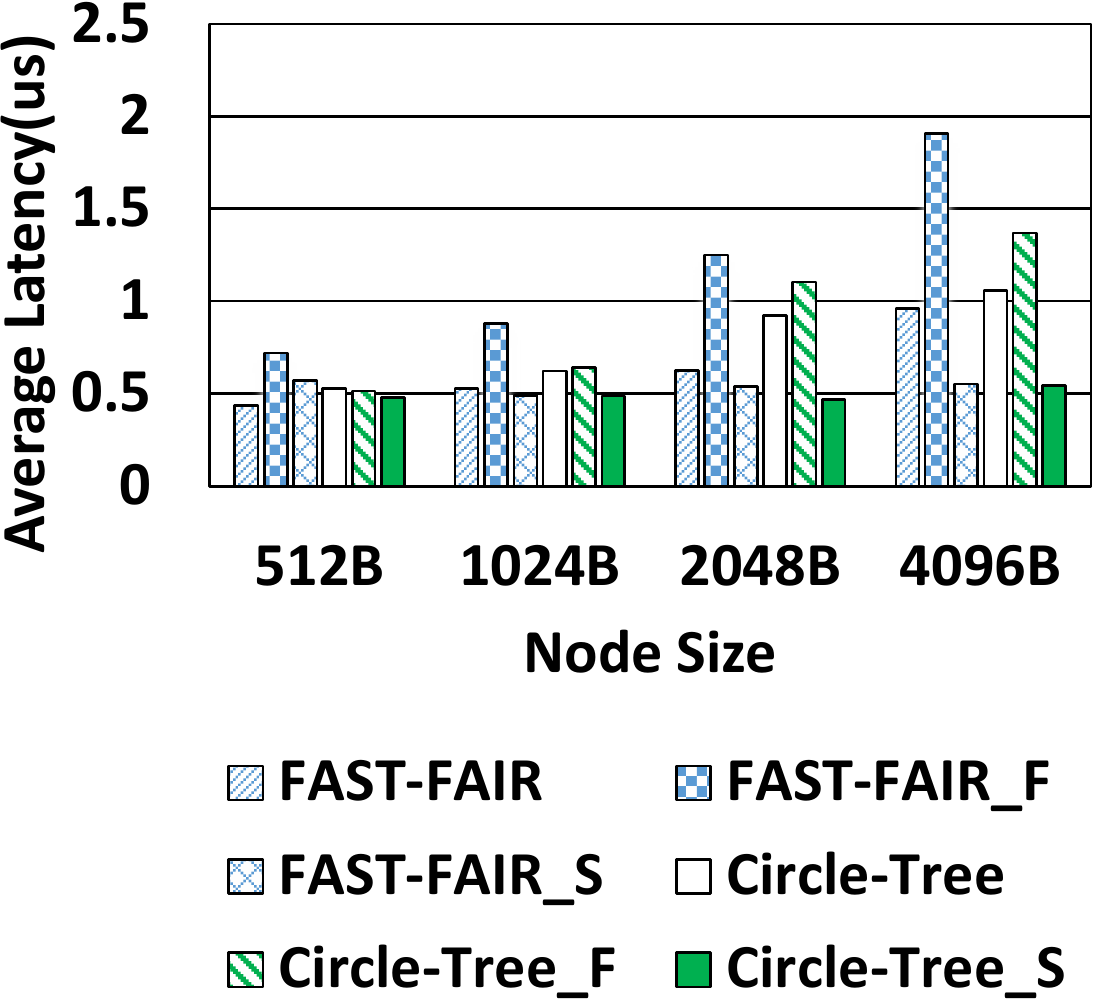}
		\label{1m_search} 
	}%
	\subfigure[Average latency (1m data, insertion, $\mu$s)]{
		\centering
		\includegraphics[width=0.48\columnwidth]{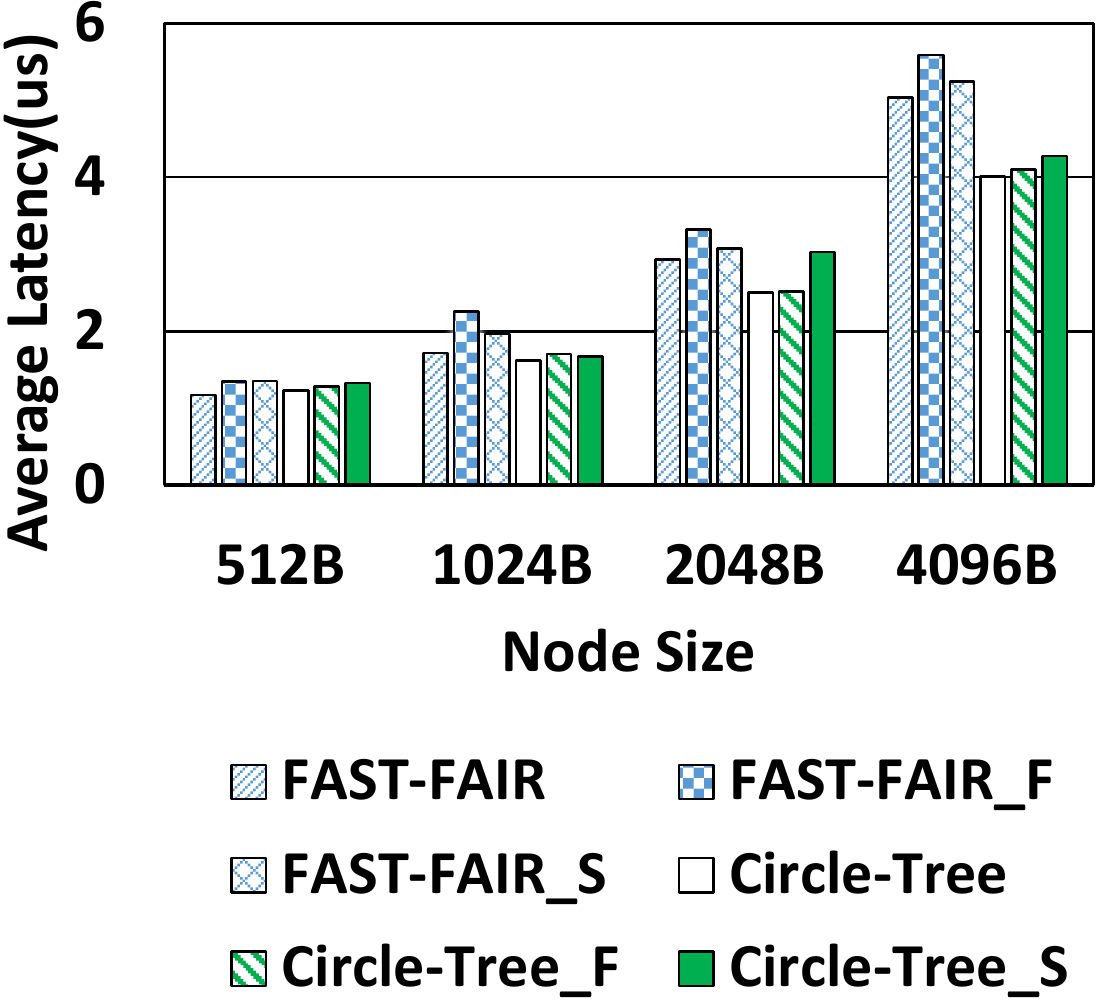}
		\label{1m_insertion} 
	}%
	\subfigure[Average latency (10m data, search, $\mu$s)]{
		\centering
		\includegraphics[width=0.48\columnwidth]{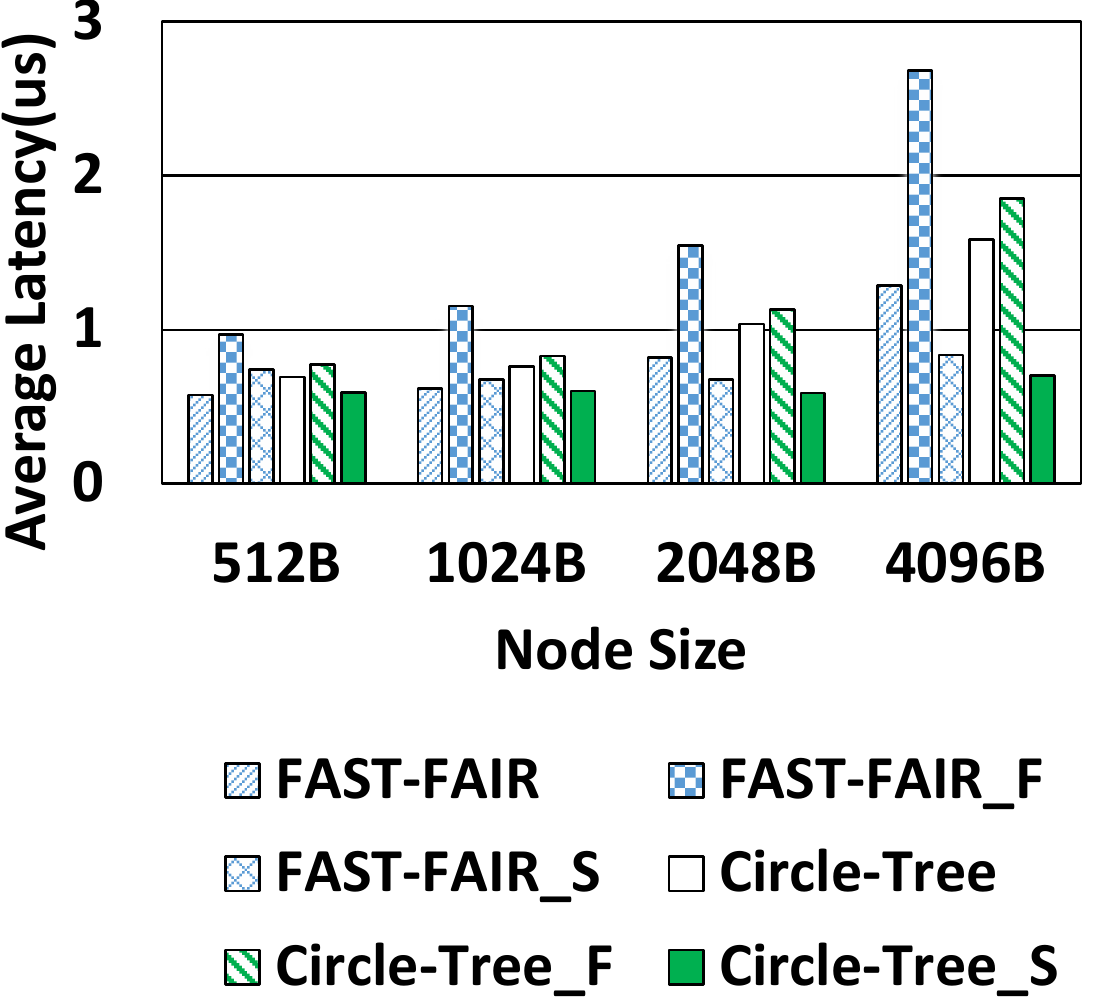}
		\label{10m_search} 
	}%
	\subfigure[Average latency (10m data, insertion, $\mu$s)]{
		\centering
		\includegraphics[width=0.48\columnwidth]{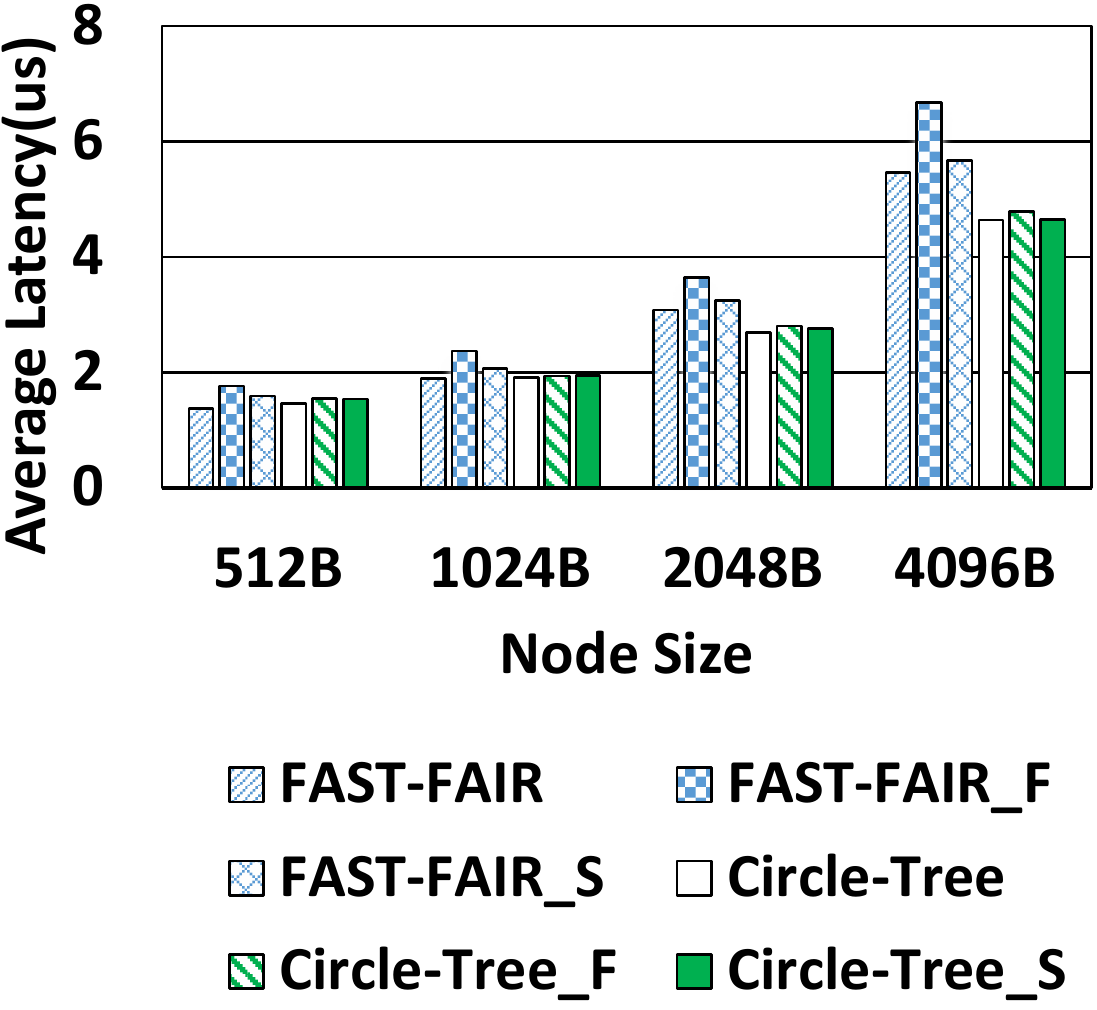}
		\label{10m_insertion} 
	}%
	\centering
	\caption{A Comparison of Six Trees on Inserting and Searching 1/10 Million Keys}
	\label{10m_testing} 
\end{figure*}

\begin{figure}[t]
	\centering
	\subfigure[Average latency (100m data, search, $\mu$s)]{
		\centering
		\includegraphics[width=0.48\columnwidth]{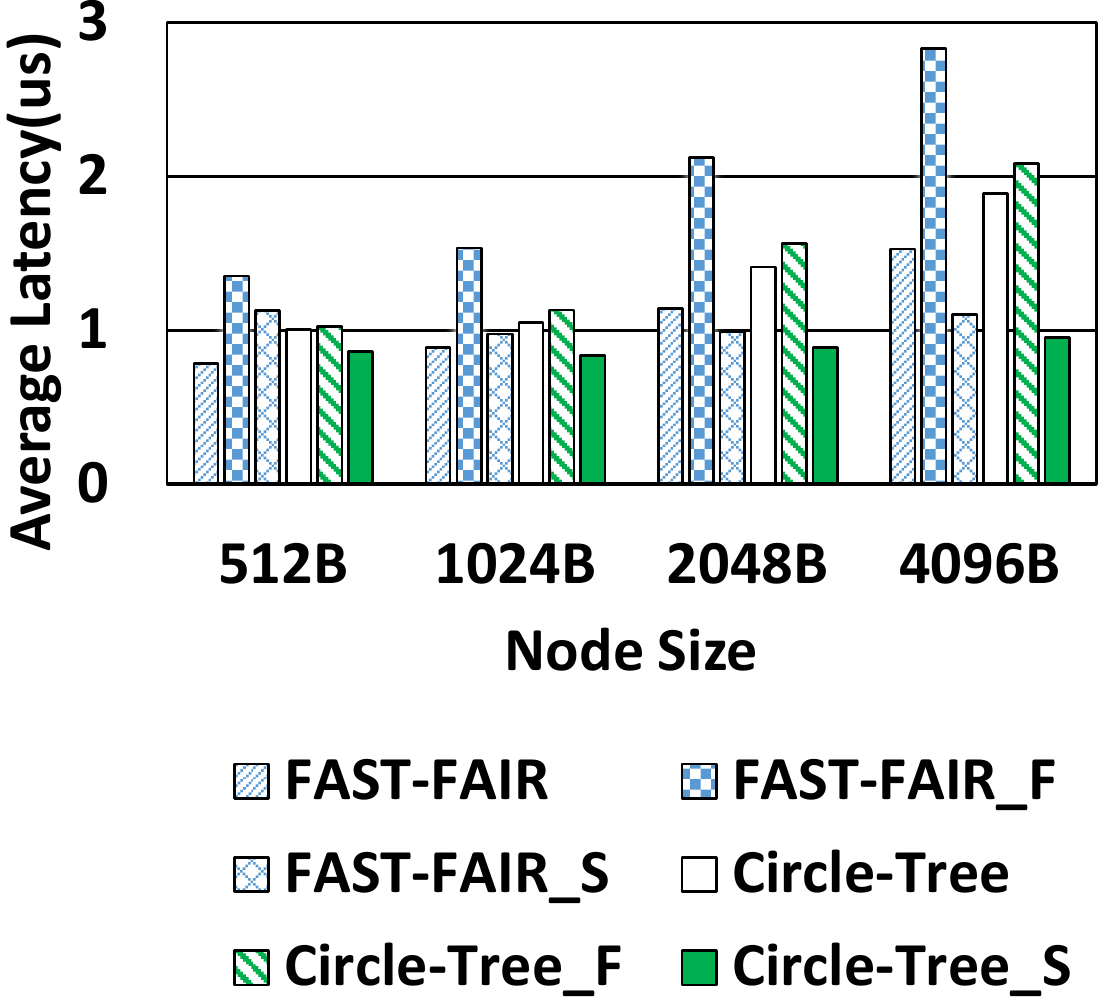}
		\label{100m_search} 
	}%
	\subfigure[Average latency (100m data, insertion, $\mu$s)]{
		\centering
		\includegraphics[width=0.48\columnwidth]{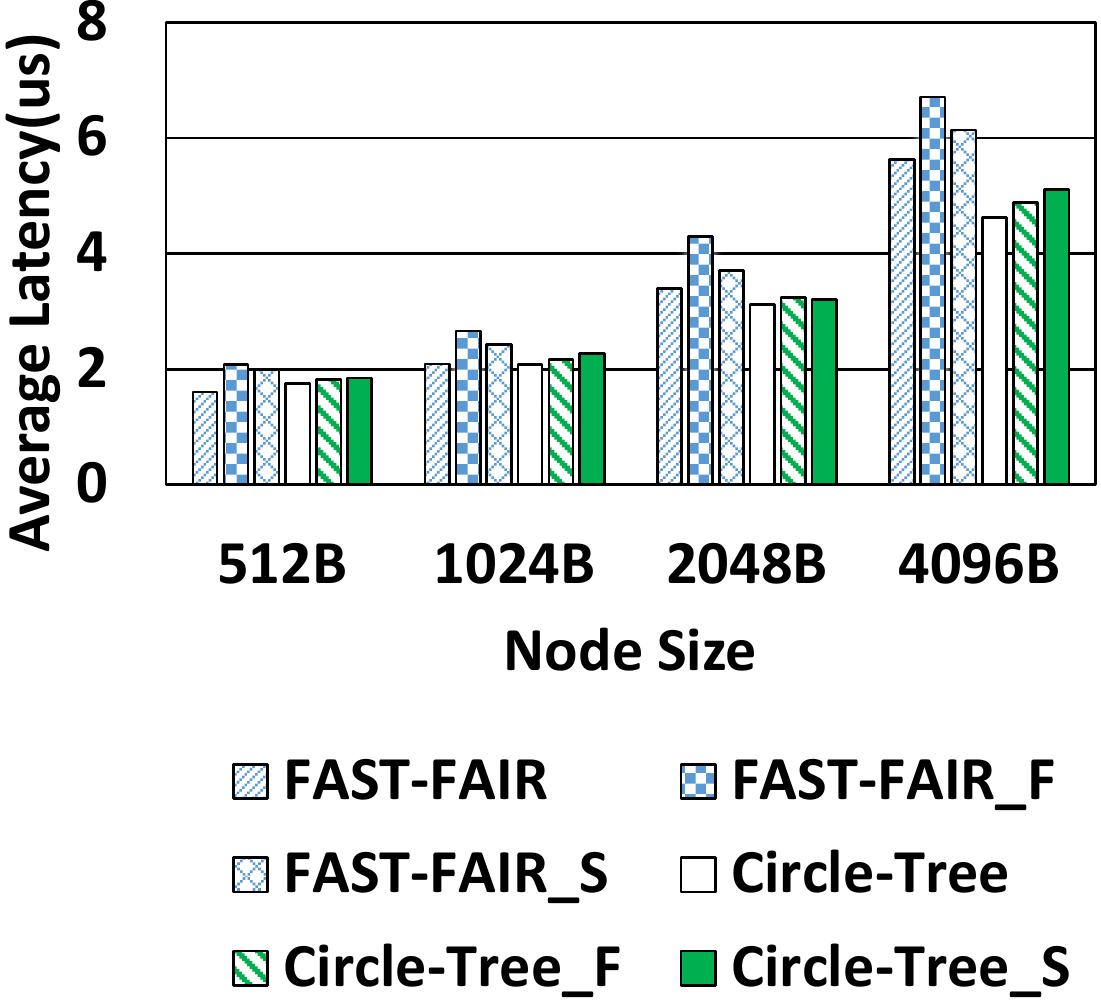}
		\label{100m_insertion} 
	}%
	\centering
	\caption{A Comparison of Six Trees on Inserting and Searching 100 Million Keys}
	\label{100m_testing} 
\end{figure}

Fig.~\ref{search_alg} exemplifies the process of searching with Sentinel Array. The target key is 140. We assume that four KV pairs occupy one cache line for 16B KV pointer size in total and 64B cache line's size.  As shown by the left part of Fig. 4, to reach 140, we need to traverse four cache lines, thereby causing four cache misses. With a Sentinel Array shown on the right part of Fig. 4, we only need to check the cache line of sentinels and the actual cache line holding key 140 and its corresponding value, i.e., two cache misses. Algorithm \ref{alg:search} illustrates the algorithmic steps of searching with Sentinel Array. We note that we add Sentinel Arrays to a B+-tree's leaf nodes without loss of generality.

\begin{algorithm}[!htbp]
    \caption{Search With Sentinel Array}
    \label{alg:search}
    \begin{algorithmic}[1]
        \REQUIRE The target key.
        \ENSURE The pointer of value of target key.
        \STATE $begin\_index = 0$
        \FOR{($i=1$; $i<\frac{n}{count\_in\_line}$; $i++$)}
            \STATE $//$ $count\_in\_line$ is number of KV pairs counted in one cache line
            \IF{$key < sentinel\_array[i]$}
                \STATE break
            \ENDIF
            \STATE $begin\_index = begin\_index + count\_in\_line$
        \ENDFOR
        \STATE Search the node's array begin with $begin\_index$
    \end{algorithmic}
\end{algorithm}

\subsection{Update of Sentinel Array}
From time to time, keys are inserted and deleted in a B+-tree node. Therefore, we need to update the node's Sentinel Array, which is simply updated with the original data array.  Because the Sentinel Array is reconstructable from consistent KV pairs, the update of it is efficient without using cache line flush or memory fence.

\subsection{Multi-threading Access}

Sentinel array is added to a B+-tree node as an auxiliary component. A node-level lock shared with the original B+-tree node helps to support multi-threading write/read with the Sentinel Array.

\section{Evaluation}

In this section, we implement and evaluate our Sentinel Array based on FAST-FAIR and Circle-Tree in terms of insertion and search performance. We use a synthetic benchmark YCSB \cite{cooper2010benchmarking} to evaluate the performance between original FAST-FAIR/Circle-Tree and ones with Sentinel Array. 

\subsection{Evaluation Setup}\label{AA}
All of our experiments are conducted on Linux Server (Kernel version 3.10) with an Intel \textsuperscript{\textregistered} Xeon \textsuperscript{\textregistered} E5-2620v4 2.10GHz CPU with 512KB/2MB/20MB L1/L2/L3 cache, 8GB DRAM. We use DRAM to emulate the NVM space. We keep the read latency of NVM as the same as DRAM, and emulate the write latency by adding an extra delay after each \textsf{clflushshopt} instruction. We set the default write latency of NVM as 300ns. \cite{yang2015nv, zhao2013kiln, wang2019circ}

\begin{figure}[t]
\centering
\subfigure[Average latency (1m data, search, $\mu$s)]{
\centering
\includegraphics[width=0.48\columnwidth]{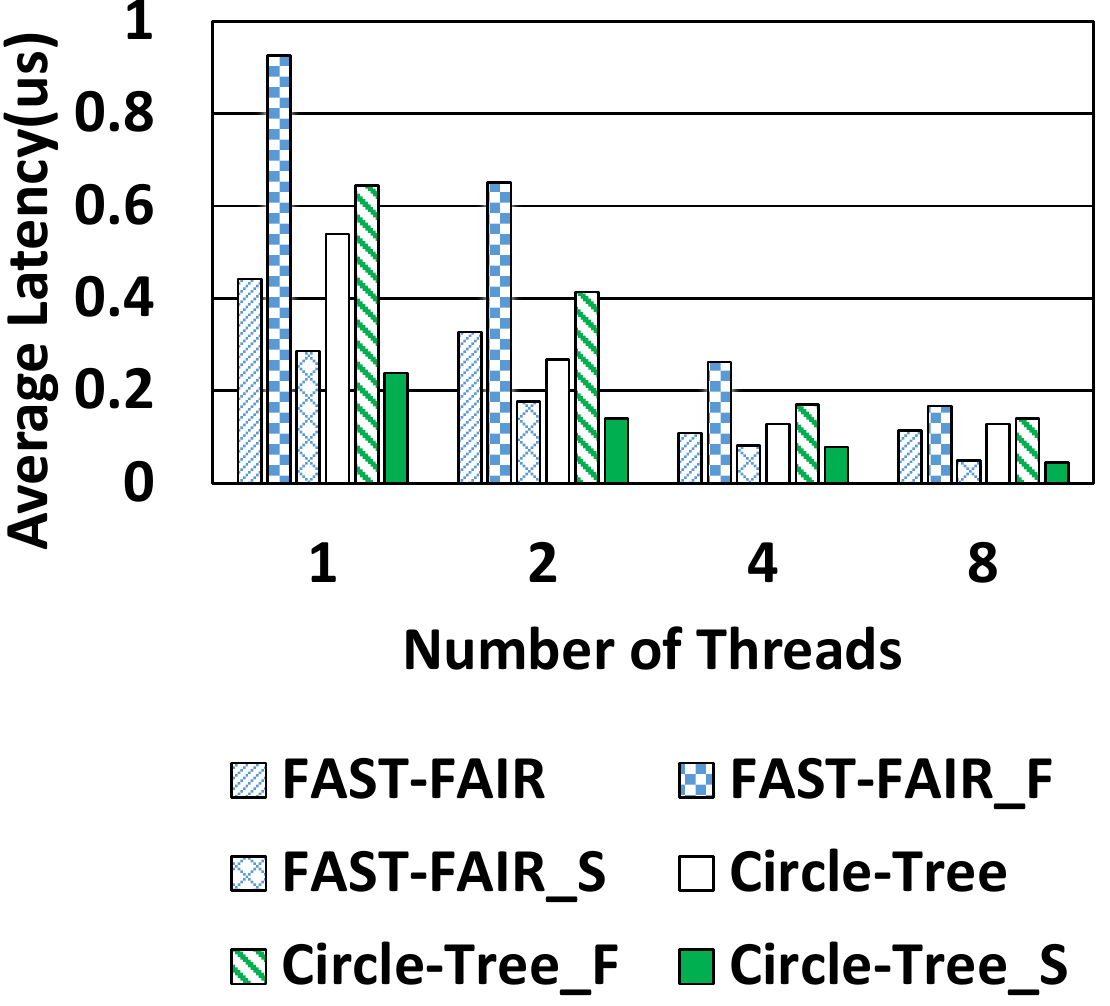}
\label{concurrent_search} 
}%
\subfigure[Average latency (1m data, insertion, $\mu$s)]{
\centering
\includegraphics[width=0.48\columnwidth]{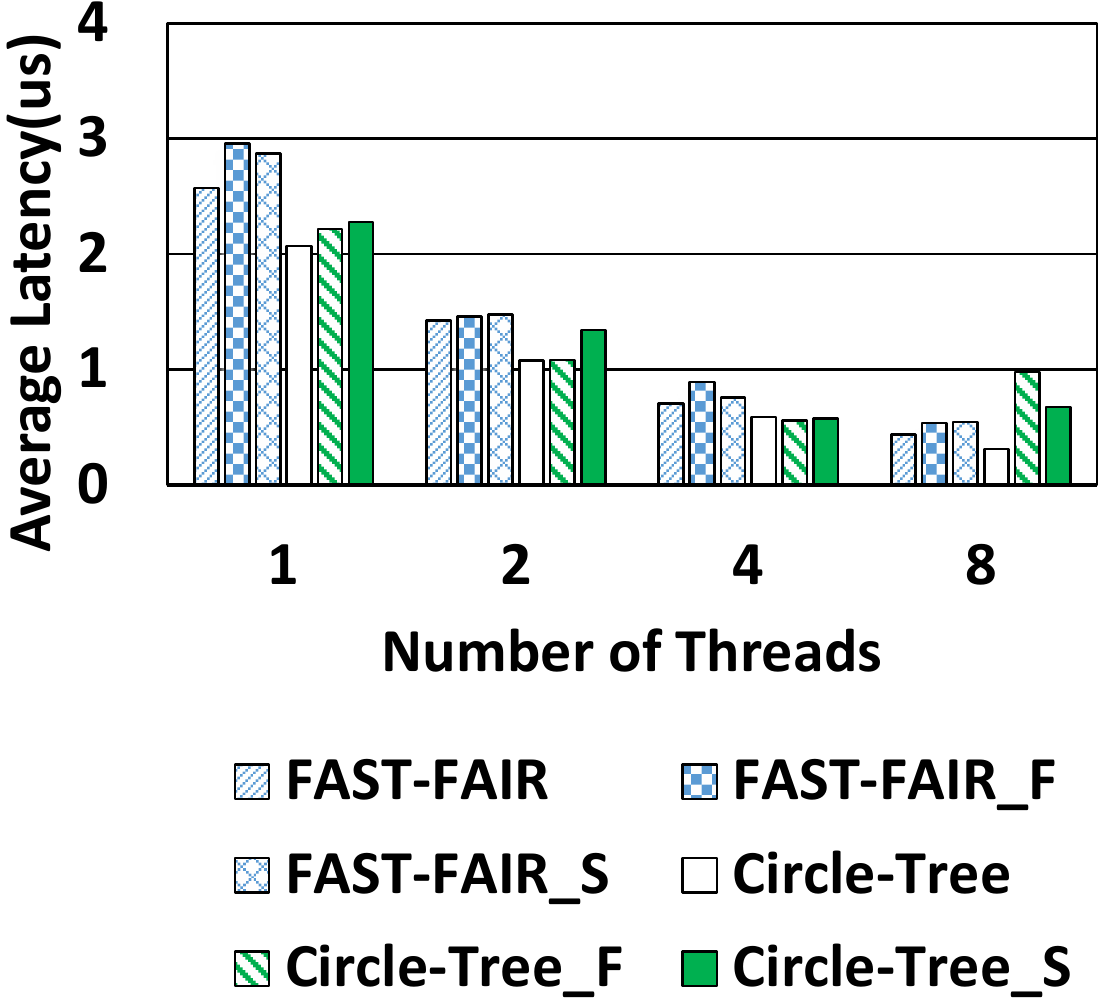}
\label{concurrent_insertion} 
}%
\centering
\caption{A Comparison of Six Trees with Multi-Threading}
\label{concurrent_testing} 
\end{figure}

\begin{figure*}[t]
\centering
\subfigure[99th percentile latency (insertion, $\mu$s)]{
\centering
\includegraphics[width=0.64\columnwidth]{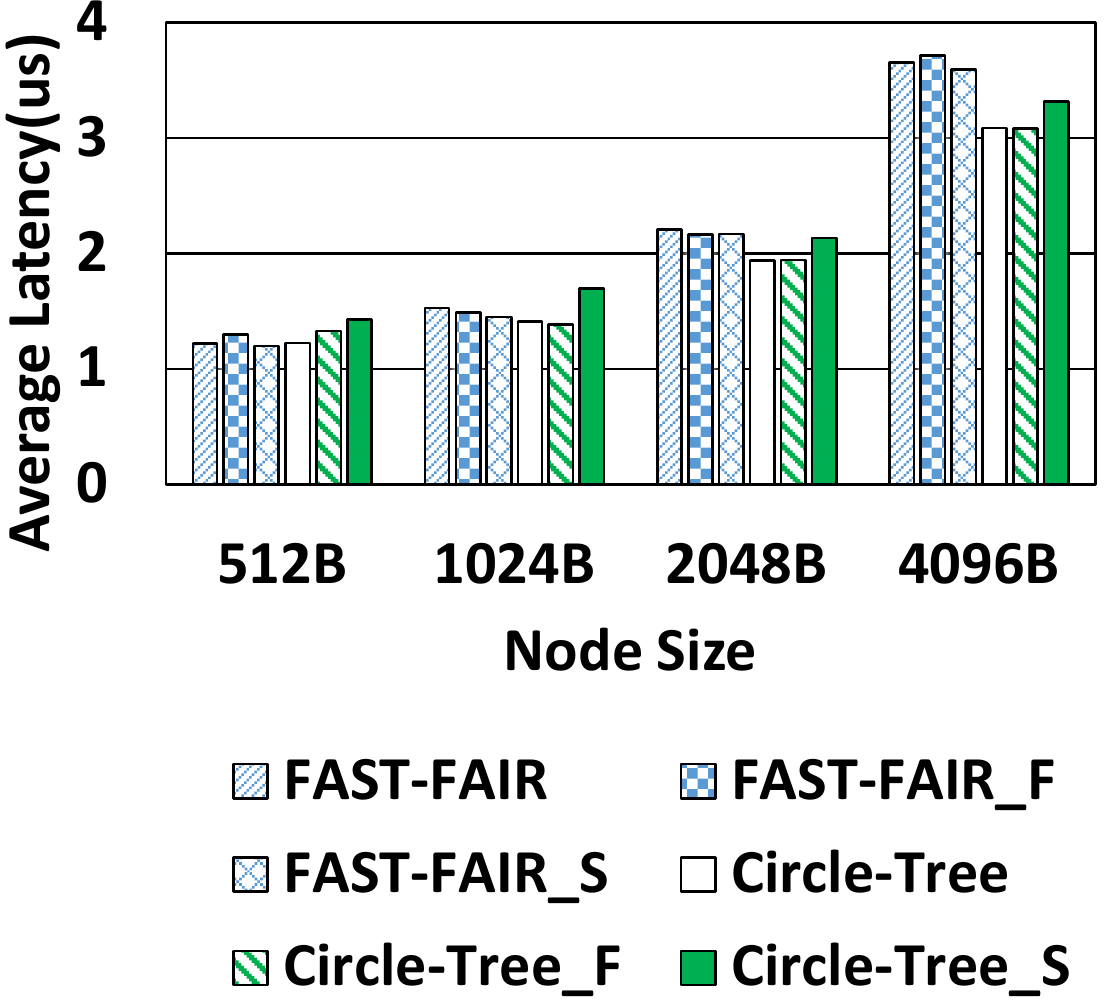}
\label{YCSB_load} 
}%
\subfigure[99th percentile latency (search, $\mu$s)]{
\centering
\includegraphics[width=0.64\columnwidth]{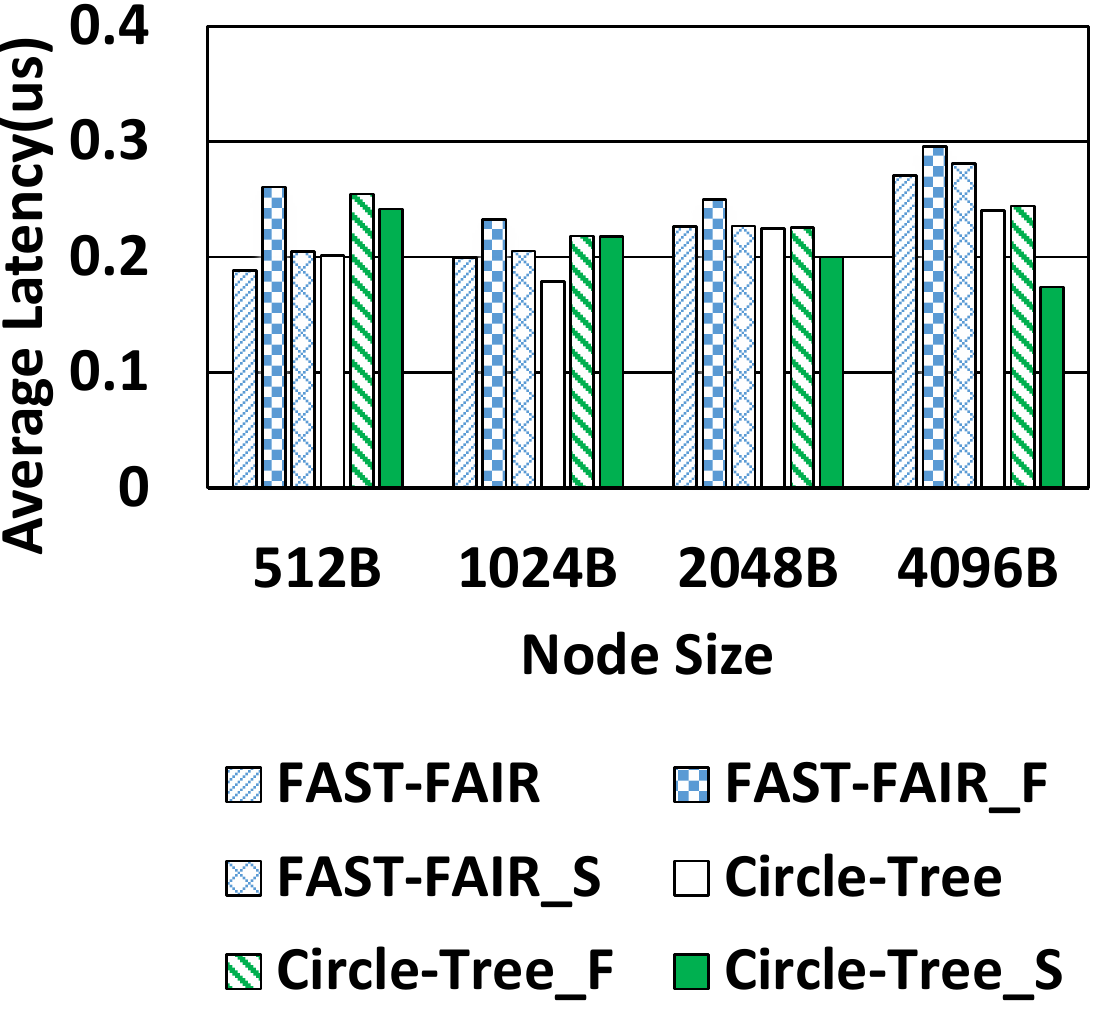}
\label{YCSB_search} 
}%
\subfigure[99th percentile latency (update, $\mu$s)]{
\centering
\includegraphics[width=0.64\columnwidth]{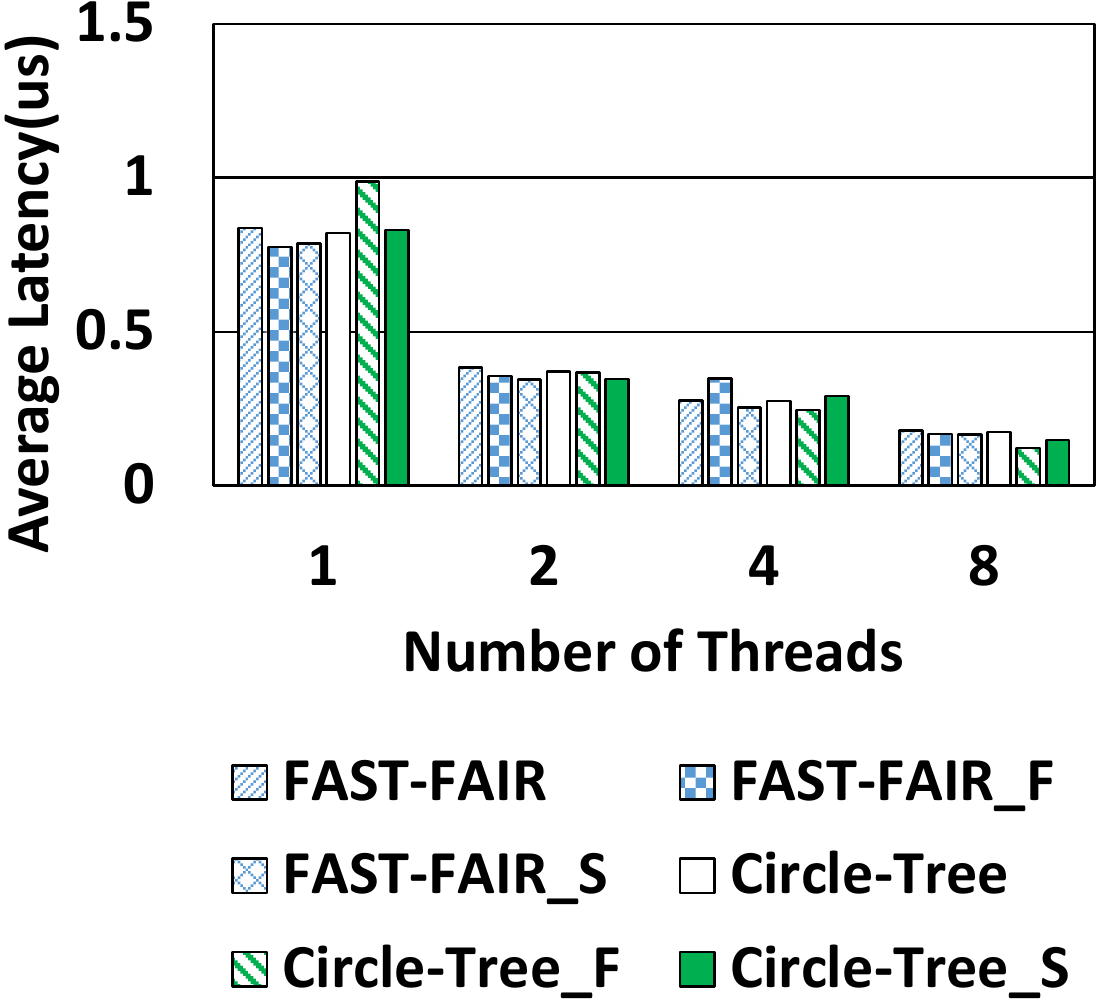}
\label{YCSB_update} 
}%
\centering
\caption{A Comparison of six KV Stores (4KB Node) on SessionStore Workload of YCSB}
\end{figure*}

\begin{figure*}[t]
\centering
\subfigure[99th percentile latency (insertion, $\mu$s)]{
\centering
\includegraphics[width=0.64\columnwidth]{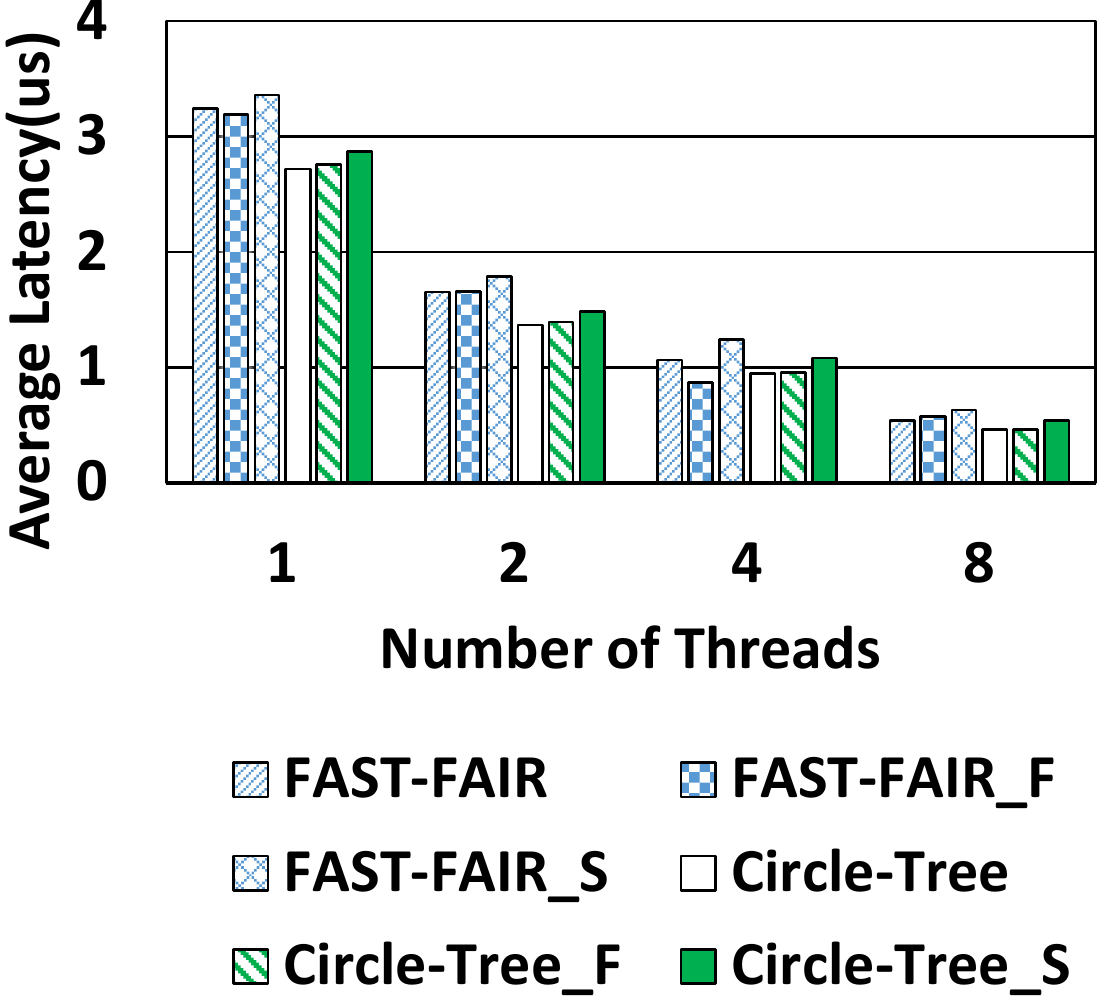}
\label{YCSB_multi_load} 
}%
\subfigure[99th percentile latency (search, $\mu$s)]{
\centering
\includegraphics[width=0.64\columnwidth]{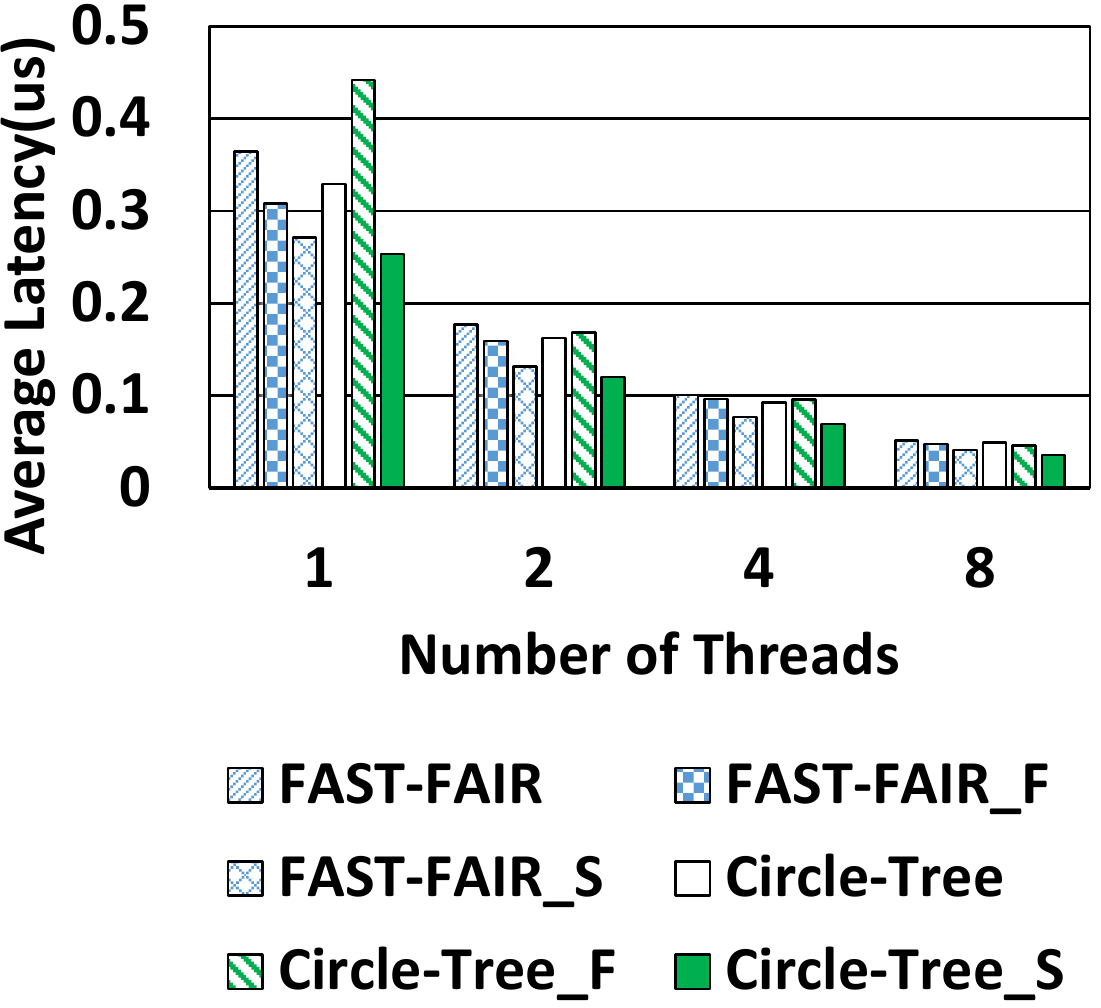}
\label{YCSB_multi_search} 
}%
\subfigure[99th percentile latency (update, $\mu$s)]{
\centering
\includegraphics[width=0.64\columnwidth]{./YCSB_multi_update.pdf}
\label{YCSB_multi_update} 
}%
\centering
\caption{A Comparison of six KV Stores (4KB Node) with Multi-threading on SessionStore Workload of YCSB}
\end{figure*}

Fig.~\ref{10m_testing} and Fig.~\ref{100m_testing} show the insertion and search performance of multi-version B+-tree variants. FAST-FAIR has open-source code and Circle-Tree is implemented with respect to its original literature. Given the large gap between the performance of the search function mentioned in the Circle-Tree and the linear search in FAST-FAIR. The Circle-Tree's search function was implemented as a linear search starting from the logical base location of a node to the last valid KV pair. Moreover, the name that has the `S' suffix is the Sentinel Array. The one that has the `F' suffix contains a fingerprint array proposed in the FP-Tree. All implementations have been compiled with -O option.

Fig.~\ref{10m_testing} and Fig.~\ref{100m_testing} show average latencies by running 1/10/100 million insertion and search operations.  A shorter average latency means higher performance. We have chosen the geometric mean to calculate the average latency. 

From Fig.~\ref{1m_search}, a B+-tree with Sentinel Array outperforms the one without Sentinel Array with much shorter search latencies. For example, using the idea of Sentinel Array helps to reduce the average search latency of Circle-Tree and FAST-FAIR by 48.4\% and 42.6\%, respectively, with 4KB node size. However, as also shown in Fig.~\ref{1m_search}, the effect of using fingerprints is limited. For example, the search latency of Circle-Tree with 4KB node size is shortened by 3.1\%. 

Although the Sentinel Array influences the insertion performance because of its update needs, still with 4KB node, the average insertion latencies of Circle-Tree and FAST-FAIR with the Sentinel Array are increased by 6.5\% and 4.0\% that of the same structure without the Sentinel Array in Fig.~\ref{1m_insertion}. 

Another observation illustrated in Fig.~\ref{10m_testing} is that the search performance of the B+-tree with the Sentinel Array gets better with larger node size. Search latency of B+-tree with Sentinel Array is larger than the ones without Sentinel Array with 512B node. The reason is that CPU combined with the compiler has branch predict feature and could prefetch the data into cache-line. The node with small size node could find the target with less cache misses, but the Sentinel Array implementation needs at least one solid cache miss to search Sentinel Array and more branch judgment between Array switch. So it's better to add the Sentinel Array to the B+-tree like data structure with large node size. The same results are also observed in Fig.~\ref{100m_testing}, which refer to more data insertion and search.

\subsection{Multi-threading Performance}

We have performed multi-threading tests with 1/2/4/8 threads, to evaluate the performance of the B+-tree with Sentinel Array. We evaluated the tests with 4KB node size and the data size is 1 million keys. Fig.~\ref{concurrent_search} shows the performance of each multi-threading search operation and we could observe that FAST-FAIR and Circle-Tree with the Sentinel Array outperform the one without the Sentinel Array. For example, using sentinels reduces the average search latency by 46.1\% and 47.6\%, for FAST-FAIR and Circle-Tree with 2 threads.

The multi-thread insertion showed in Fig.~\ref{concurrent_insertion}, indicates that the Sentinel Array has insignificant influence on insertion performance.

\subsection{Evaluation on YCSB}

For end-to-end comparison, we built an interface for each B+-tree like data structure to receive and handle access requests issued by YCSB. Because the key from a YCSB workload is a string with a prefix and a number, we removed the prefix and treated the number as an 8B key in unsigned integer. The default load value of YCSB has ten fields with 100 bytes per field, so the pointer of a KV pair in a leaf node points to the address of stored value.

We used the `workloada' workload with YCSB. It first inserts a predefined number of KV pairs ({\em load}) and then follows a search/update ratio of 50\%/50\% over keys selected in accordance with a Zipf distribution ({\em run}). YCSB reports a series of latencies from which we choose the 99th percentile latency to rule out the impact of very short or very long operations. It means that 99\% of overall write/read requests can be completed below such a latency.

In our experiment, we use 1 million KV pairs to test performance with YCSB. Fig.~\ref{YCSB_load} shows the average latencies of inserting 1 million data into each tree. The structure with the Sentinel Array still gets lower latency than the original implementation. Fig.~\ref{YCSB_search} captures search performance and FAST-FAIR and Circle-Tree with the Sentinel Array get 16.9\% and 13.0\% improvement on 4KB node size. From Fig.~\ref{YCSB_update}, which represents the average latency of updating the KV pair of the loaded data, the structure with the Sentinel Array still gets a shorter latency. The reason is that update operation is like the search operation and they only need to find the KV pair's position without shifting the KV pair.

We also tested multi-threads of the B+-tree with Sentinel Array with YCSB. As showed in Fig.~\ref{YCSB_multi_load}, Fig.~\ref{YCSB_multi_search} and Fig.~\ref{YCSB_multi_update} the average search latencies is significantly reduced compared to ones without Sentinel Array.
\section{Conclusion}
Search is of primary importance for B+-tree. In this paper, we revisited the search operation of in-NVM B+-tree with sorted nodes and found that the search performance is mainly affected by read amplification and cache misses. As a result, we proposed to use a sentinel to stand for each cache line of a sorted node. Before checking the actual array of KV pairs, a search operation first goes through the probing Sentinel Array to determine which cache line the target key would reside in. As the Sentinel Array is much smaller than the actual array of KV pairs, the number of cache misses is significantly reduced. Extensive experiments confirm that using sentinels significantly boosts the search performance of state-of-the-art B+-trees developed for NVM, especially with larger nodes. 

{
\bibliographystyle{IEEEtran}	
\bibliography{sentinel}

\begin{thebibliography}{10}
\providecommand{\url}[1]{#1}
\csname url@samestyle\endcsname
\providecommand{\newblock}{\relax}
\providecommand{\bibinfo}[2]{#2}
\providecommand{\BIBentrySTDinterwordspacing}{\spaceskip=0pt\relax}
\providecommand{\BIBentryALTinterwordstretchfactor}{4}
\providecommand{\BIBentryALTinterwordspacing}{\spaceskip=\fontdimen2\font plus
\BIBentryALTinterwordstretchfactor\fontdimen3\font minus
  \fontdimen4\font\relax}
\providecommand{\BIBforeignlanguage}[2]{{%
\expandafter\ifx\csname l@#1\endcsname\relax
\typeout{** WARNING: IEEEtran.bst: No hyphenation pattern has been}%
\typeout{** loaded for the language `#1'. Using the pattern for}%
\typeout{** the default language instead.}%
\else
\language=\csname l@#1\endcsname
\fi
#2}}
\providecommand{\BIBdecl}{\relax}
\BIBdecl

\bibitem{b1}
L.~M. Grupp, J.~D. Davis, and S.~Swanson, ``The bleak future of {NAND} flash
  memory,'' in \emph{Proceedings of the 10th USENIX Conference on File and
  Storage Technologies}, ser. FAST’12.\hskip 1em plus 0.5em minus 0.4em\relax
  USA: USENIX Association, 2012, p.~2.

\bibitem{lee2009architecting}
\BIBentryALTinterwordspacing
B.~C. Lee, E.~Ipek, O.~Mutlu, and D.~Burger, ``Architecting phase change memory
  as a scalable dram alternative,'' vol.~37, no.~3.\hskip 1em plus 0.5em minus
  0.4em\relax New York, NY, USA: Association for Computing Machinery, Jun.
  2009, p. 2–13. [Online]. Available:
  \url{https://doi.org/10.1145/1555815.1555758}
\BIBentrySTDinterwordspacing

\bibitem{qureshi2009scalable}
\BIBentryALTinterwordspacing
M.~K. Qureshi, V.~Srinivasan, and J.~A. Rivers, ``Scalable high performance
  main memory system using phase-change memory technology,'' in
  \emph{Proceedings of the 36th Annual International Symposium on Computer
  Architecture}, ser. ISCA ’09.\hskip 1em plus 0.5em minus 0.4em\relax New
  York, NY, USA: Association for Computing Machinery, 2009, p. 24–33.
  [Online]. Available: \url{https://doi.org/10.1145/1555754.1555760}
\BIBentrySTDinterwordspacing

\bibitem{zhou2009durable}
\BIBentryALTinterwordspacing
P.~Zhou, B.~Zhao, J.~Yang, and Y.~Zhang, ``A durable and energy efficient main
  memory using phase change memory technology,'' p. 14–23, 2009. [Online].
  Available: \url{https://doi.org/10.1145/1555754.1555759}
\BIBentrySTDinterwordspacing

\bibitem{raoux2008phase}
S.~Raoux, G.~W. Burr, M.~J. Breitwisch, C.~T. Rettner, Y.-C. Chen, R.~M.
  Shelby, M.~Salinga, D.~Krebs, S.-H. Chen, H.-L. Lung \emph{et~al.},
  ``Phase-change random access memory: A scalable technology,'' \emph{IBM
  Journal of Research and Development}, vol.~52, no. 4.5, pp. 465--479, 2008.

\bibitem{kawahara2010scalable}
T.~Kawahara, ``Scalable spin-transfer torque {RAM} technology for normally-off
  computing,'' \emph{IEEE Design Test of Computers}, vol.~28, no.~1, pp.
  52--63, 2011.

\bibitem{chang2016resistance}
T.-C. Chang, K.-C. Chang, T.-M. Tsai, T.-J. Chu, and S.~M. Sze, ``Resistance
  random access memory,'' \emph{Materials Today}, vol.~19, no.~5, pp. 254--264,
  2016.

\bibitem{bayer2002organization}
\BIBentryALTinterwordspacing
R.~Bayer and E.~M. Mccreight, ``Organization and maintenance of large ordered
  indexes.''\hskip 1em plus 0.5em minus 0.4em\relax Berlin, Heidelberg:
  Springer-Verlag, Sep. 1972, vol.~1, no.~3, p. 173–189. [Online]. Available:
  \url{https://doi.org/10.1007/BF00288683}
\BIBentrySTDinterwordspacing

\bibitem{yang2015nv}
\BIBentryALTinterwordspacing
J.~Yang, Q.~Wei, C.~Chen, C.~Wang, K.~L. Yong, and B.~He, ``{NV-Tree}: Reducing
  consistency cost for {NVM-based} single level systems,'' in \emph{13th
  {USENIX} Conference on File and Storage Technologies ({FAST} 15)}.\hskip 1em
  plus 0.5em minus 0.4em\relax Santa Clara, CA: {USENIX} Association, Feb.
  2015, pp. 167--181. [Online]. Available:
  \url{https://www.usenix.org/conference/fast15/technical-sessions/presentation/yang}
\BIBentrySTDinterwordspacing

\bibitem{hwang2018endurable}
\BIBentryALTinterwordspacing
D.~Hwang, W.-H. Kim, Y.~Won, and B.~Nam, ``Endurable transient inconsistency in
  byte-addressable persistent {B+-Tree},'' in \emph{16th {USENIX} Conference on
  File and Storage Technologies ({FAST} 18)}.\hskip 1em plus 0.5em minus
  0.4em\relax Oakland, CA: {USENIX} Association, Feb. 2018, pp. 187--200.
  [Online]. Available:
  \url{https://www.usenix.org/conference/fast18/presentation/hwang}
\BIBentrySTDinterwordspacing

\bibitem{wang2019circ}
C.~Wang, G.~Brihadiswarn, X.~Jiang, and S.~Chattopadhyay, ``Circ-tree: A
  {B+-Tree} variant with circular design for persistent memory,'' \emph{arXiv
  preprint arXiv:1912.09783}, 2019.

\bibitem{moraru2013consistent}
\BIBentryALTinterwordspacing
I.~Moraru, D.~G. Andersen, M.~Kaminsky, N.~Tolia, P.~Ranganathan, and
  N.~Binkert, ``Consistent, durable, and safe memory management for
  byte-addressable non volatile main memory,'' in \emph{Proceedings of the
  First ACM SIGOPS Conference on Timely Results in Operating Systems}, ser.
  TRIOS ’13.\hskip 1em plus 0.5em minus 0.4em\relax New York, NY, USA:
  Association for Computing Machinery, 2013. [Online]. Available:
  \url{https://doi.org/10.1145/2524211.2524216}
\BIBentrySTDinterwordspacing

\bibitem{venkataraman2011consistent}
S.~Venkataraman, N.~Tolia, P.~Ranganathan, and R.~H. Campbell, ``Consistent and
  durable data structures for non-volatile byte-addressable memory,'' in
  \emph{Proceedings of the 9th USENIX Conference on File and Stroage
  Technologies}, ser. FAST’11.\hskip 1em plus 0.5em minus 0.4em\relax USA:
  USENIX Association, 2011, p.~5.

\bibitem{oukid2016fptree}
I.~Oukid, J.~Lasperas, A.~Nica, T.~Willhalm, and W.~Lehner, ``{FPTree}: A
  hybrid {SCM-DRAM} persistent and concurrent {B-tree} for storage class
  memory,'' in \emph{Proceedings of the 2016 International Conference on
  Management of Data}, 2016, pp. 371--386.

\bibitem{zhao2013kiln}
\BIBentryALTinterwordspacing
J.~Zhao, S.~Li, D.~H. Yoon, Y.~Xie, and N.~P. Jouppi, ``Kiln: Closing the
  performance gap between systems with and without persistence support,'' in
  \emph{Proceedings of the 46th Annual IEEE/ACM International Symposium on
  Microarchitecture}, ser. MICRO-46.\hskip 1em plus 0.5em minus 0.4em\relax New
  York, NY, USA: Association for Computing Machinery, 2013, p. 421–432.
  [Online]. Available: \url{https://doi.org/10.1145/2540708.2540744}
\BIBentrySTDinterwordspacing

\bibitem{mohan1992aries}
C.~Mohan, D.~Haderle, B.~Lindsay, H.~Pirahesh, and P.~Schwarz, ``{ARIES}: a
  transaction recovery method supporting fine-granularity locking and partial
  rollbacks using write-ahead logging,'' \emph{ACM Transactions on Database
  Systems (TODS)}, vol.~17, no.~1, pp. 94--162, 1992.

\bibitem{dulloor2014system}
\BIBentryALTinterwordspacing
S.~R. Dulloor, S.~Kumar, A.~Keshavamurthy, P.~Lantz, D.~Reddy, R.~Sankaran, and
  J.~Jackson, ``System software for persistent memory,'' in \emph{Proceedings
  of the Ninth European Conference on Computer Systems}, ser. EuroSys
  ’14.\hskip 1em plus 0.5em minus 0.4em\relax New York, NY, USA: Association
  for Computing Machinery, 2014. [Online]. Available:
  \url{https://doi.org/10.1145/2592798.2592814}
\BIBentrySTDinterwordspacing

\bibitem{intel64and}
Intel, ``Intel{\textregistered} 64 and {IA-32} architectures software
  developer’s manual,'' \emph{Volume 3A: System Programming Guide, Part},
  vol.~1, no.~64, p.~64, Sep. 2016.

\bibitem{lu2014loose}
Y.~Lu, J.~Shu, L.~Sun, and O.~Mutlu, ``Loose-ordering consistency for
  persistent memory,'' in \emph{2014 IEEE 32nd International Conference on
  Computer Design (ICCD)}, 2014, pp. 216--223.

\bibitem{danowitz2012cpu}
A.~Danowitz, K.~Kelley, J.~Mao, J.~P. Stevenson, and M.~Horowitz, ``{CPU DB}:
  recording microprocessor history,'' \emph{Queue}, vol.~10, no.~4, pp. 10--27,
  2012.

\bibitem{cooper2010benchmarking}
B.~F. Cooper, A.~Silberstein, E.~Tam, R.~Ramakrishnan, and R.~Sears,
  ``Benchmarking cloud serving systems with {YCSB},'' in \emph{Proceedings of
  the 1st ACM symposium on Cloud computing}, 2010, pp. 143--154.

\end{thebibliography}
}

\end{document}